\documentclass[twocolumn,showpacs,preprintnumbers,amsmath,amssymb,superscriptaddress]{revtex4}

\usepackage{graphicx}
\usepackage{dcolumn}
\usepackage{bm}
\usepackage{graphicx}
\usepackage{xspace}
\usepackage{multirow}
\usepackage{natbib}
\usepackage{color}

\renewcommand{\vec}{\mathbf}

\newcommand{\rcco}{RE$_{2-x}$Ce$_x$CuO$_4$}

\newcommand{\pcco}{Pr$_{2-x}$Ce$_x$CuO$_4$}

\newcommand{\plcco}{Pr$_{1-x}$LaCe$_{x}$CuO$_4$}
\newcommand{\rlcco}{RE$_{1-x}$LaCe$_{x}$CuO$_4$}

\newcommand{\nccox}[2]{Nd$_{#1}$Ce$_{#2}$CuO$_4$}
\newcommand{\pccox}[2]{Pr$_{#1}$Ce$_{#2}$CuO$_4$}

\newcommand{\plccox}[2]{Pr$_{#1}$LaCe$_{0.12}$CuO$_4$}

\begin{document}

\preprint{}

\title{Antiferromagnetism-superconductivity competition in electron-doped cuprates triggered by oxygen reduction}

\author{P. Richard}\email{richarpi@bc.edu}
\affiliation{Department of Physics, Boston College, Chestnut Hill, MA 02467}
\author{M. Neupane}
\affiliation{Department of Physics, Boston College, Chestnut Hill, MA 02467}
\author{Y.-M. Xu}
\affiliation{Department of Physics, Boston College, Chestnut Hill, MA 02467}
\author{P. Fournier}
\affiliation{D\'{e}partement de physique, Universit\'{e} de Sherbrooke, Sherbrooke, Qu\'{e}bec, Canada, J1K 2R1}
\author{S. Li}
\affiliation{Department of Physics and astronomy, The University of Tennessee, Knoxville, TN 37996}
\author{Pengcheng Dai}
\affiliation{Department of Physics and astronomy, The University of Tennessee, Knoxville, TN 37996}
\affiliation{Neutron Scattering Sciences Division, Oak Ridge National Laboratory, Oak Ridge, TN 37831}
\author{Z. Wang}
\affiliation{Department of Physics, Boston College, Chestnut Hill, MA 02467}
\author{H. Ding}
\affiliation{Department of Physics, Boston College, Chestnut Hill, MA 02467}

\date{\today}

\begin{abstract}
We have performed a systematic angle-resolved photoemission study of as-grown and oxygen-reduced \pcco\xspace 
and \plcco\xspace electron-doped cuprates. In contrast to the common belief, neither the band filling nor the band parameters are significantly 
affected by the oxygen reduction process. Instead, we show that the main electronic role of the reduction process is to remove an 
anisotropic leading edge gap around the Fermi surface. While the nodal leading edge gap is induced by long-range antiferomagnetic order, the origin of the antinodal one remains unclear. 
\end{abstract}


\maketitle


Even though most of the work has been focused on hole-doped cuprates, the understanding 
of their electron-doped counterparts is essential for obtaining a universal picture of high-Tc superconductivity. To achieve this goal, it is necessary to first solve the main mystery that holds since the discovery 
of the T'-structure electron-doped cuprates \rcco\xspace and  \rlcco\xspace (RE = Pr, Nd, Sm, Eu): why is superconductivity in these 
compounds achieved only when a tiny amount of oxygen ($\sim1\%$)  is removed from the as-grown (AG) samples following a 
post-annealing process (reduction) \cite{Moran,Kim,Wang,Takayama,Susuki}? In fact, the AG samples, even with sufficient 
electron-doping by adding Ce, are antiferromagnetic (AF) insulators at low temperature. Far from being a simple materials issue, 
the understanding of the microscopic origin of the reduction process that triggers superconductivity may shed light on 
other related questions in high-Tc superconductivity. 

Long considered as the microscopic explanation of the reduction process, the removal of extraneous oxygen atoms located above 
Cu (apical oxygen) has been ruled out by recent Raman and crystal-field infrared transmission studies \cite{Riou,Richard1}. 
Indeed, these studies revealed two main defects appearing with the oxygen reduction, which have been tentatively assigned to out-of-plane and in-plane oxygen vacancies, the latter being the only one observed at optimal doping. In parallel, a (RE,Ce)$_2$O$_3$ impurity phase epitaxial to the CuO$_2$ planes appears in reduced superconducting samples but disappears in re-oxygenated nonsuperconducting samples \cite{Kurahashi,Matsuura,Dai1,Kang3}. 
Based on this phenomenon, it has been proposed recently that the Cu excess released during the formation of the (RE,Ce)$_2$O$_3$ impurity phase 
fills Cu vacancies and makes the remaining structure more stoichiometric \cite{Kang3}. Which of these 
structural defects has the most significant impact on the electronic properties is still under intense debate and calls for a better 
characterization of the electronic band structure before and after the reduction process. In contrast to the widespread belief that the reduction 
process in the electron-doped cuprates can be considered as an independent degree of freedom for carrier doping, a recent systematic study of 
the Hall coefficient in \pcco\xspace thin films with various oxygen contents showed that the carrier mobility rather than their 
concentration is modified by the reduction process \cite{Gauthier}. In particular, the annealing process leads to the delocalization of 
holelike carriers, most likely due to the suppression of the long-range AF order \cite{Richard4,Dai1}. Understanding how the reduction process 
can tune the competition between the AF and superconducting states and modify the electronic band structure is thus crucial. 

In this letter, we present the first systematic angle-resolved photoemission spectroscopy (ARPES) study of the impact of the reduction process in the electron-doped cuprates. We show that 
neither the electronic filling nor the band structure parameters are significantly changed by the reduction process. Instead, the reduction process suppresses long-range AF order and fills up a leading edge gap (LEG) which has two components. While the nodal LEG is of AF origin, the nature of the antinodal LEG remains unclear.


High-quality \pccox{1.85}{0.15}\xspace  single crystals have been grown by the flux technique. Some of the nonsuperconducting AG
 samples have been annealed in an argon environment at temperatures between 850 and 925 $^o$C for a typical period of 5 days 
encapsulated in a polycrystalline matrix \cite{Brinkmann} and are referred in the text as reduced samples. The reduced samples exhibit superconducting transitions 
 around 24 K. Using the floating zone technique, high-quality \plccox{0.88}{0.12}\xspace single crystals have also been grown and have been annealed as described in Ref. \cite{Kang3}. The samples have been studied by ARPES using the PGM 
 and U1-NIM beamlines of the Synchrotron Radiation Center  (Stoughton, WI) with photon energies of 73.5 and 22 eV. The data have been 
 recorded at 40 K using a Scienta SES-2002 analyzer with a 30 meV energy resolution. The samples have been cleaved \emph{in situ} 
 in a vacuum better than 10$^{-10}$ Torr. Although this letter focuses on the data obtained on \pcco, similar results have been obtained for 
 the \plcco\xspace samples.  

\begin{figure}[htbp]
\begin{center}
\includegraphics[width=9cm]{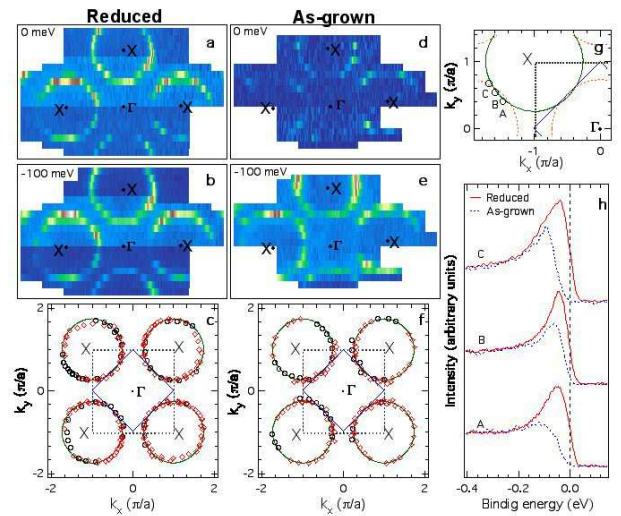}
\caption{\label{fig_fs}(Color online) a-b and d-e) CEIPs (20 meV integration) of the ARPES data obtained at 40 K using a 73.5 eV photons 
with a $\vec{A}||\Gamma$-X polarization. c and f) Underlying FS contours associated with the reduced and AG samples. The experimental data are represented by circles while 
the points indicated by diamonds have been obtained by symmetry operations. The data have been fitted by an effective tight-binding model 
(see the text). h) Comparison of the EDCs of reduced (solid) and AG (dashed) at the locations given in panel g.} 
\end{center}
\end{figure}

In order to transform AG samples into superconductors, the reduction process must affect the electronic structure and especially the 
excitations near the Fermi energy (E$_{\textrm{F}}$). Fig. \ref{fig_fs} compares the constant energy intensity plots (CEIPs) of the 
reduced (a,b) and AG (d,e) samples in momentum space, as obtained by ARPES. Bright spots indicate regions with large 
photoemission intensity. The CEIPs centered at -100 meV with 20 meV energy integration window (b,e) are quite similar, and one can easily distinguish the 
X($\pm\pi$,$\pm\pi$)-centered holelike pockets. This contrasts with the CEIPs centered at E$_{\textrm{F}}$ (a,d). Contrary to the reduced 
sample (Fig. \ref{fig_fs}a), the intensity at E$_{\textrm{F}}$ is strongly suppressed in the AG sample (Fig. \ref{fig_fs}d). Nevertheless, 
underlying Fermi surface (FS) contours can be extracted in both cases and the results are reproduced in Figs. \ref{fig_fs}c and f for the 
reduced and AG samples, respectively. Surprisingly, the data extracted for the reduced and the AG samples can be fitted, 
within uncertainties, with the same band parameters $\mu=0.05$ eV, $t=-1.1$ eV and $t^{\prime}=0.32$ eV, using the simple effective tight-binding (TB)
model $E-\mu=t/2[\cos(k_x)+\cos(k_y)]+t^{\prime}\cos(k_x)\cos(k_y)$. 

According to Luttinger theorem, the introduction of extra negative 
carriers following the reduction would lead to smaller X-centered holelike pockets and thus to at least an increase of $\mu$, which is not 
observed. Actually, the underlying FS contours coincide with a doping of x $\approx$ 0.15 in both cases. This is a strong evidence that the 
reduction process modifies neither the band filling nor the shape of the band dispersion sufficiently to induce the dramatic changes 
observed in the transport properties \cite{Gauthier}. Instead, the CEIPs at E$_\textrm{F}$ indicate that the annealing process removes a LEG that is present at E$_\textrm{F}$ in the AG samples. This assertion is confirmed by Fig. \ref{fig_fs}h, which compares 
the electron distribution curves (EDCs) of AG and reduced samples at different k-locations given in Fig. \ref{fig_fs}g. All the AG sample spectra have their leading edge shifted towards higher binding energies as compared with the corresponding reduced sample spectra and thus have much weaker intensities at E$_\textrm{F}$. 

\begin{figure}[htbp]
\begin{center}
\includegraphics[width=8cm]{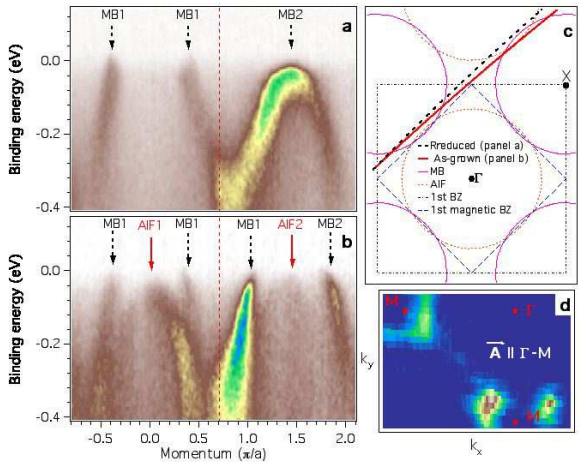}
\caption{\label{long_c}(Color online) Comparison of the reduced (a) and AG (b) spectra obtained along the lines given in c. MB (dashed arrows) and AIF (solid arrows) 
indicate the main and AIF band structures, respectively. The number next to MB or AFI indicates the zone in which the band is 
detected. d) CEIP at -50 meV (30 meV integration) of a \pccox{1.85}{0.15}\xspace AG sample obtained at 22 eV. The suppression along the vertical $\Gamma$-M direction is due to ARPES selection rules.}
\end{center}
\end{figure}

Now one asks the question: how can the reduction process suppress the LEG observed in the AG samples? We first 
checked that the samples were not charged by increasing the photon flux, which had no influence on the leading edge shift in our experiment. 
The most likely candidate to explain this mystery is the AF ordering, which exists in AG samples. It is known that AF is suppressed in the 
reduced samples \cite{Richard4,Dai1}. This effect is clearly observable by ARPES. We plotted in Figs. \ref{long_c}a and \ref{long_c}b the 
electronic dispersion as measured along the lines indicated in Fig. \ref{long_c}c for the reduced (dashed) and AG (solid) samples, 
respectively. In addition to the main band branches, indicated with dashed arrows, the spectrum of the AG sample shows features that are
 not observed in the reduced one. As indicated directly by solid arrows on Fig. \ref{long_c}, these features correspond to the AF-induced folding 
 (AIF) band. Such features, observed in the first as well as in the second Brillouin zones (BZs), are responsible for the M(-$\pi$,0)-centered electron 
 FS pockets emphasized in Fig. \ref{long_c}d, which shows the AG CEIP at -50 meV obtained with 22 eV photons. 

In the presence of an AIF band, the hybridization of the main and AIF bands opens a gap at locations coinciding with the magnetic BZ 
boundary \cite{Matsui}. The energy position of the center of the gap between the upper and lower hybridized bands depends on the k-location along that 
boundary, defined by M(0,$\pi$) and equivalent points. Hence, along the (0,$\pi$)$\rightarrow$($\pi$,0) direction, it occurs below 
E$_\textrm{F}$ between the M points and the hot spots, which are defined as the k-locations where the intersection occurs exactly at E$_\textrm{F}$. On the other hand, between the hot spots, the intersection occurs always above E$_\textrm{F}$ and the upper hybridized band can never 
cross E$_\textrm{F}$ and therefore cannot be observed by low temperature ARPES, whereas the top of the lower hybridized band is pushed down. 
When the AF gap is large enough, the small holelike FS pocket around ($\pi$/2,$\pi$/2) is gapped out. 

\begin{figure}[htbp]
\begin{center}
\includegraphics[width=8cm]{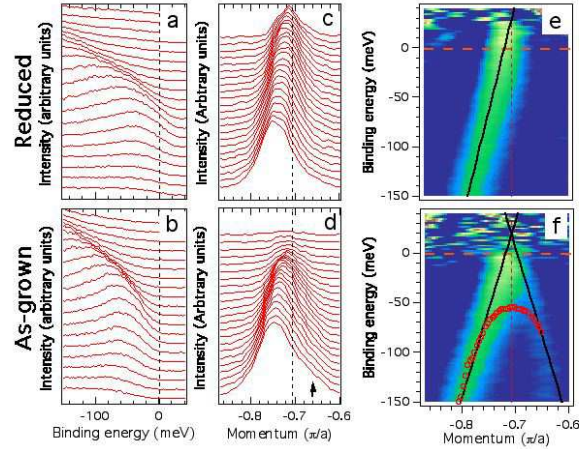}
\caption{\label{fig_nodal}(Color online) Comparison of the nodal dispersion (second zone) between reduced (top panels) and AG samples 
(bottom panels). a-b) EDCs from -0.87$\pi$/a (bottom) to -0.62$\pi$/a (top). c-d) MDCs between - 80 meV (bottom) and 0 meV (top).  e-f) SMDI plots. The vertical dashed line corresponds to the AF boundary, while the solid lines and the dots correspond to the unhybridized main and AF bands, and to the hybridized band, respectively.}
\end{center}
\end{figure}

In order to check this scenario, we investigated the band dispersion of reduced and AG samples along the nodal ($\Gamma$-X) direction. 
The results are given in Fig. \ref{fig_nodal}. Figs. \ref{fig_nodal}a and b show the EDCs of the reduced and AG samples, respectively. 
Besides the clear leading edge shift observed for the AG sample as compared to the reduced one, the  EDC maxima of the AG 
sample exhibit a bending back characteristic of hybridization: from the top to the bottom of Fig. \ref{fig_nodal}b, the EDC maxima first move 
closer to E$_\textrm{F}$, and then move away, with a decrease of intensity. 

A contrast in the shape of  the momentum distribution curves (MDCs) 
of the reduced and AG samples, which are given respectively  in Figs. \ref{fig_nodal}c and d, is also observed. The asymmetric shape of 
the AG MDCs suggests the presence of a band folded along the AF boundary (vertical dashed line). This effect is clearly seen on the corresponding second momentum-derivative intensity (SMDI) plots displayed in Figs.  \ref{fig_nodal}e and f for the reduced and AG 
samples, respectively. The position of the MDC peaks corresponds to the minimum in the SMDI plots (bright spots). In contrast to the situation shown in Fig. \ref{fig_nodal}e, Fig. \ref{fig_nodal}f exhibits an additional 
band whose dispersion is the reflection of the main band with respect to the AF boundary. Using the position of the MDCs, we extracted the 
main band dispersion and reported it on Fig. \ref{fig_nodal}f, along with its reflection across the AF boundary. We also reported on 
Fig. \ref{fig_nodal}f the position of the EDC maxima associated with the AG sample. These maxima, which  coincide with the renormalized 
dispersion band, support the hybridization scenario and indicate that the portion of the FS around ($\pi$/2,$\pi$/2) is suppressed in the 
AG samples. 
  
  
\begin{figure}[htbp]
\begin{center}
\includegraphics[width=8cm]{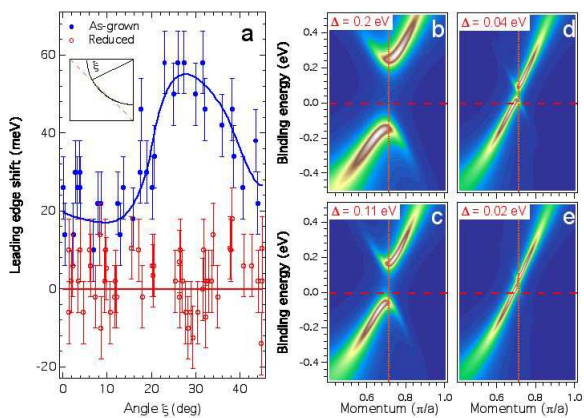}
\caption{\label{kdep}(Color online) a) k-dependece of the leading edge shift. Lines are guides for the eye. b-e) Simulations of the band dispersion in the presence of an AF gap. 
We used the TB parameters defined in the text and introduced a band broadening in order to mimic 
real data.}
\end{center}
\end{figure}

Fig. \ref{kdep}a, which compares the k-dependence of the leading edge shift for the AG and reduced samples, provides additional evidence 
that an AF hybridization gap is suppressed after the reduction process. While no clear leading edge shift is observable for the reduced sample, 
an anisotropic LEG is observed for the AG sample. Hence, a maximum is observed around the hot spot, as expected from 
the hybridization scenario \cite{Matsui}. In order to illustrate further the AF scenario, we plotted in Figs. \ref{kdep}b-e simulations of the nodal dispersion obtained using 
the fit parameters given above, with various AF gap sizes. We introduced a broadening to mimic realistic results and removed the 
Fermi function for sake of clarity. For a large gap, the band folding is clearly observable and the lower hybridized band never crosses 
E$_{\textrm{F}}$. As a consequence, a large leading edge shift is recorded. This leading edge shift decreases as the gap becomes smaller, 
and it disappears when the lower band crosses E$_{\textrm{F}}$, as illustrated in \ref{kdep}d. 
Our simulations indicate that a LEG of 20 meV along the nodal direction can be produced by a 60 meV AF LEG at the hot spot with a proper linewidth, in agreement with our observation in Fig. \ref{kdep}a.
 
While the experimental results for the nodal region can easily be described by an AF hybridization gap, such a model alone 
fails to explain the LEG observed for the antinodal region, where the main and AIF bands intersect below 
E$_{\textrm{F}}$. In particular, the original band position at M is $\sim$ 300 meV below E$_{\textrm{F}}$, thus a small LEG ($\sim$ 20 meV) cannot be produced by a simple AF hybridization effect. Nevertheless, AF order may still be responsible for the antinodal LEG. It has been predicted that the whole FS of the \nccox{1.85}{0.15}\xspace superconducting samples, including the antinodal region, is pseudogapped due to paramagnons in the semiclassical regime \cite{Sunko}. Even though this is in apparent contradiction with our experimental data, which indicate no antinodal LEG for the reduced samples, this idea may be valid for the AG samples, for which the AF correlations are much stronger. 

We now turn to a critical question: how can a small amount of extra oxygen atoms induce AF order in the AG samples? Literature provides two opposite scenarios involving CuO$_2$ plane defects and the competition between the AF and superconducting states. It has been suggested that in-plane oxygen vacancies in the reduced samples can suppress the AF order, affect the band parameters and induce superconductivity  \cite{Riou,Richard1}. However, the present study indicates that the band parameters are not modified significantly by the reduction process. Moreover, the antinodal LEG in this scenario would be more likely an indirect consequence of magnetic fluctuations such as paramagnons \cite{Sunko}. The lack of  theoretical study on the subject leaves open the possibility that charge disorder induced by oxygen vacancies can suppress the AF order and promote superconductivity.

An opposite scenario is based on a recent neutron study suggesting a deficiency of Cu in AG \plcco\xspace samples that is healed after the reduction process through the formation of a (RE,Ce)$_2$O$_3$ impurity phase \cite{Kang3}. It has been argued that, in hole-doped cuprates, a Cu vacancy, like a nonmagnetic Zn impurity, would suppress local superconducting phase coherence and at the same time induce a staggered paramagnetic S=1/2 local moment extending over a few unit cells \cite{Z_Wang}. Such impurity or Cu vacancy induced local magnetism has been recently observed by in-plane $^{17}$O NMR in the superconducting state of the optimally hole-doped YBa$_2$Cu$_3$O$_7$ with dilute Zn impurities \cite{Ouazi}. If the amount of Cu vacancies in the AG samples, although small, is sufficient, it is possible to establish AF long-range order by quantum percolation of the AF regions. In addition, the strong scattering of the Cu vacancies in CuO$_2$ planes of AG samples may produce a localization gap (or Coulomb gap) that could explain the observed antinodal LEG. We note that the residual resistivity ($\sim$ 500 $\mu\Omega$cm) at the superconductor-insulator transition, 
found in re-oxygenated \pccox{1.83}{0.17}\xspace thin films \cite{Gauthier}, 
corresponds to the two-dimensional (2D) resistance $\rho_{0}^{2D} \approx$ 8.3 k$\Omega/\square$ per CuO$_2$, close to the 
universal 2D value $h/4e^2 \simeq 6.5 k\Omega/\square$ \cite{Pang,Fisher,Emery}. Similar results were also found in the Zn-substituted
hole-doped cuprates \cite{Fukuzumi}. Moreover, the 2\%-4\% of Zn substitution needed to suppress completely superconductivity in 
La$_{2-x}$Sr$_x$Cu$_{1-z}$Zn$_z$O$_4$  \cite{Fukuzumi} is similar to the value of 1.2\% to 2.3\% Cu vacancies estimated in the AG 
and non-superconducting \plccox{0.88}{0.12}\xspace samples \cite{Kang3}. 

We caution that there are other possible explanations to account for the antinodal LEG, such as a charge density wave induced by the nesting of two sides of the M-centered electron pockets. However, it would then be hard to explain why both the long-range AF and this charge density wave are suppressed in the reduced samples. The unexpected presence of the antinodal LEG calls for further theoretical and experimental studies. 

We are indebted to A.-M. S. Tremblay and D. S\'{e}n\'{e}chal for useful discussions. We acknowledge support from NSF DMR-0353108, DOE DEFG02-99ER45747 and DE-FG02-05ER46202. This work is based upon research conducted at the Synchrotron Radiation Center supported by NSF DMR-0537588. P.F. acknowledges the support of NSERC (Canada), FQRNT (Qu\'{e}bec), CFI and CIAR.

\bibliography{biblio_ens}

\end{document}